# New Bounds on Spherical Antenna Bandwidth and Directivity: Updates to the Chu-Harrington Limits

Carl Pfeiffer and Bae-Ian Wu

*Abstract* — The Chu circuit model provides the basis for analyzing the minimum radiation quality factor, $Q$, of a given spherical mode. However, examples of electrically large spherical radiators readily demonstrate that this $Q$ limit has limitations in predicting bandwidth. Spherical mode radiation is reexamined and an equivalent 1D transmission line model is derived that exactly models the fields. This model leads to a precise cutoff frequency of the spherical waveguide, which provides a clear boundary between propagating and evanescent fields. A new delineation of 'stored' and 'radiated' electromagnetic energy is postulated, which leads to a new definition of spherical mode $Q$. Next, attention is turned to the Harrington bound on the directivity-bandwidth tradeoff of an antenna with an arbitrary size. Harrington derived the maximum directivity for a specified number of spherical harmonics such that the $Q$ is not 'large'. Here, the method of Lagrange multipliers is used to quantify the maximum directivity for a given bandwidth. It is shown that optimally exciting all spherical harmonics (including $n > ka$) enables both larger directivity and bandwidth than Harrington's previous limit. While Chu and Harrington's analyses are generally good approximations for most situations, the new self-consistent theory that defines fundamental antenna limits leads to updated results.

## I. Introduction

Much of antenna theory is devoted towards understanding the available trade space between size, bandwidth, efficiency, and gain of antennas. The impact of material loss on radiation efficiency and gain is studied in [1, 2, 3, 4, 5]. Our work will instead focus on the relationships between size, bandwidth, and directivity, of a lossless antenna. The maximum available bandwidth is a complicated function of the specifics of the antenna and the complexity of the matching network as shown by Bode and Fano [6, 7]. Therefore, the quality factor ($Q$) is typically studied to understand general frequency characteristics of an antenna. The traditional relationship between half-power impedance bandwidth, $B_{3dB}$, and $Q$ given by $B_{3dB} = 2/Q$ is rigorously valid when $Q \gg 1$ and there is a single resonance. For multi-resonant designs and $Q \gg 1$, the inverse relationship between $Q$ and bandwidth is approximate. When $Q$ is on the order of unity or less there is even more

The authors are with Air Force Research Laboratory, Wright-Patterson Air Force Base, OH 45433 (e-mail: carlpfei@umich.edu, bae-ian.wu.1@us.af.mil)

ambiguity. In this regime, it is only possible to state that broad bandwidth behavior on the order of an octave or more is expected. The terms $Q$ and bandwidth will be used interchangeably here subject to the above-mentioned limitations. The $Q$ of an antenna with purely resistive input impedance is generally defined as the ratio of the energy 'stored' in the electric and magnetic fields ($W_s$) to the power radiated ($P$),

$$Q = \frac{\omega W_s}{P} \qquad (1)$$

where $\omega$ is the angular frequency. However, the delineation between 'stored' vs. 'radiated' energy is often ambiguous.

In 1945 Chu derived a simple equivalent circuit that perfectly models the wave impedance of an arbitrary spherical mode at all frequencies [8]. He postulated that energy stored in the reactive circuit elements corresponds to energy stored in the outward propagating wave, which led to the most ubiquitous definition of the minimum antenna $Q$. Twenty years later, Collin and Rothschild used a field integration approach and subtracted the portion of the energy associated with radiation to define stored energy in their definition of $Q$ [9]. Collin and Rothschild arrived at the same values for $Q$ as Chu, which solidified this definition [10]. Today, Chu's circuit model for evaluating antenna $Q$ is considered to be the most rigorous with the sole limitation being it is a loose bound (i.e., overly optimistic) because the circuit only models fields external to a spherical region of space. A myriad of work has expanded on Chu's theory to develop tighter bounds on the $Q$ that accounts for energy stored within the antenna or non-spherical geometries [11, 12, 2, 13, 14]. Vandenbosch proposed a particularly notable definition for $Q$ that is commonly used to analyze arbitrarily shaped antennas [15, 16]. However, it is known that Vandenbosch's definition sometimes results in negative stored energy for electrically large structures, which is unphysical [17]. Time-domain based definitions of $Q$ have also been proposed [18, 19, 20]. A thorough review of the advantages and disadvantages of various definitions of $Q$ is reported in [21].

Curiously, using the Chu circuit model to calculate the $Q$ of high order spherical modes leads to contradictions. It will be shown that there are scenarios with simultaneously large $Q$ (i.e., narrowband) and wide impedance bandwidth. This contradiction brings into question the circuit model's validity for lower order modes, as well as all subsequent work that relies on Chu's result.

One such consideration is the directivity and bandwidth limits for antennas with arbitrary electrical sizes [22, 23, 24]. In [25], Harrington uses spherical modes to show that antennas with directivity, $D$, satisfying $D > (ka)^2 + 2ka$ must have a 'high' $Q$, where $k$ is the free space wavenumber and $a$ is the radius of the minimum sized sphere that circumscribes the antenna. However, inserting Chu's definition for $Q$ into Harrington's analysis suggests that antennas with near 100% aperture efficiency must have vanishingly small bandwidths as the size increases. This conclusion is clearly incorrect and provides further evidence that Chu's definition of $Q$ needs updating. Furthermore, Harrington stops short of quantifying the optimal tradeoff between $Q$ and directivity. A directivity-bandwidth tradeoff is quantified by Fante and Geyi, who calculate the spherical mode excitation that maximizes the ratio $D/Q$ [26, 27]. However, their optimization for $D/Q$ tends to find an excitation that achieves $Q \ll 1$, at which point there is a tenuous relationship between $Q$ and bandwidth. It will be shown that using the ratio $D/Q$ alone to determine the maximum directivity of an antenna with $Q > 1$ results in a very loose directivity bound such that it is practically useless when the antenna has a moderate electrical size (e.g., $ka > 2$). Rather than maximizing the $D/Q$ ratio, it is more useful to calculate the maximum possible directivity for a specified bandwidth.

In this paper, we introduce new bounds on the bandwidth and directivity of antennas. Like Chu and Harrington, we consider an ideal hypothetical spherical antenna that does not store any internal energy and thus realizes the optimal performance for antennas confined to a spherical volume. This also implies the bounds presented here are loose for non-spherical antennas. First, we review some of the seminal works that established fundamental limits for antennas. Example scenarios are used to clearly illustrate how these analyses have self-contradictions, which suggests these theories are only accurate in a limited sense. Arguments explaining why these previous analyses sometimes fail are also provided. Then, a new definition of the $Q$ of spherical mode radiation based on a transmission line model is proposed. In contrast to previous analyses, this new definition of $Q$ seems to accurately delineate stored and radiated energy for all electrical sizes. This new $Q$ definition is consistent with the Chu circuit model when $ka$ is small enough, which suggests that most of the previous work in electrically small antenna theory remains valid when $ka \ll 1$. However, updating Chu's definition of $Q$ for arbitrary order spherical modes is essential for analyzing antennas with moderate to large electrical sizes (e.g., $ka > 1$). For example, we study the optimal tradeoff between directivity and $Q$ for arbitrarily sized antennas. The method of Lagrange multipliers is used to calculate the optimal spherical mode excitation that maximizes directivity for a specified $Q$ (or minimum $Q$ for a specified directivity). Harrington's definition of maximum practical directivity with broad bandwidth remains approximately correct since it generally produces a directivity that is within 1.5 dB of optimal. However, it is useful to finally rigorously show it with the updated analysis. It should be noted that the discussion of the optimal spherical mode excitation is very

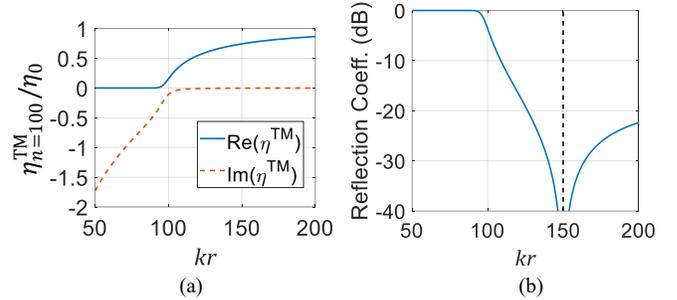

Fig. 1 (a) TM polarized wave impedance of the $n = 100$ order mode normalized by the free space wave impedance, $\eta_0$. (b) Reflection coefficient when the wave impedance is matched at $kr = 150$ with an ideal series inductor.

similar to the recent results reported in [28, 29]. The main distinction here is that we use our new $Q$ definition rather than Chu's $Q$ definition, which leads to modified results.

## II. Issues with Previous $Q$ Bounds

It is helpful to quickly review issues with some of the seminal work that has established commonly used definitions of antenna $Q$. There exists a large body of subsequent research that builds upon these foundational papers. However, the subsequent work is generally consistent with the work that is reviewed here, such that it also generally suffers from the same self-inconsistencies.

### A. Chu

In [8], Chu derives an equivalent circuit model that has the exact same wave impedance as a spherical mode. Others expanded on Chu's theory to develop tighter bounds on the $Q$ that accounts for energy stored within the antenna or electrically small non-spherical geometries [11, 12, 2, 13, 14]. Chu postulates that energy stored and dissipated in the circuit model has a one-to-one relationship with energy stored/radiated by electric and magnetic fields. However, such an equivalent circuit representation for an impedance is not unique. For example, using a transmission line to connect the radiation resistance to the rest of the circuit gives an identical input impedance. As pointed out by Kuester [30], replacing the transmission line with an equivalent LC ladder network causes the energy stored in reactive elements to increase proportionally to the length of the transmission line. In other words, two different circuit models can provide an identical input impedance and vastly different quality factors, which contradicts the notion that impedance bandwidth and $Q$ must be inversely proportional. Clearly, the energy stored in the reactive transmission line LC ladder network should be interpreted as radiated rather than stored in this example. However, this distinction between radiated and stored energy in a reactive circuit network is not generally obvious.

Consider radiation from the transverse magnetic spherical mode of order $n = 100$ (i.e., TM$_{100,m}$) around the region where $kr = 150$, where $r$ is the radius. The real and imaginary parts of the input impedance vs. $kr$ are shown in Fig. 1(a). Around $kr = 150$, the reactance is practically 0, which suggests there should be broadband properties for the mode. Fig. 1(b) plots the reflection coefficient when a small series inductance cancels the wave impedance reactance to create a broadband resonance at $kr = 150$. However, evaluating the energy stored and dissipated in reactive and resistive elements using the Chu circuit model suggests that $Q = 39$ (i.e., only a 5% half power bandwidth). Clearly, Chu's theory is not self-consistent because $Q$ has no relation to the impedance bandwidth. Further inspection of the TM$_{100,m}$ Chu circuit model reveals that when the circuit is operated above cutoff, it is well represented by the dual of a conventional transmission line (series capacitors and shunt inductors). In this regime, most of the energy stored in this transmission line should be regarded as radiated rather than stored energy. Again, there is not a clear distinction between stored and radiated energy in the reactive circuit elements, which leads to ambiguity in calculating $Q$.

It is also worth noting that Chu's $Q$ definition is most commonly applied for electrically small radiators with $ka \ll 1$ such that the reactive elements are clearly in cutoff. There is less ambiguity in defining stored and radiated energy in this regime than when $ka > n$. This could explain why the self-inconsistency of Chu's definition of $Q$ has not been reported to date.

In [8], Chu also introduces a mathematically convenient approximate formula for calculating $Q$ based on modelling the spherical mode wave impedance as a series RLC circuit. It is interesting that this approximate analysis provides more reasonable values of $Q$ for $ka > n$ than the more rigorous definition based on Chu's ladder network. However, there is no physical justification as to why replacing Chu's ladder circuit network with a series RLC circuit provides a more accurate estimate for antenna $Q$. In fact, the series RLC circuit approximation for calculating $Q$ is only valid when the input resistance is approximately constant (i.e., $d[\text{Re}(Z_{\text{in}})]/d\omega \ll d[\text{Im}(Z_{\text{in}})]/d\omega$). This relation is not generally true when $Z_{\text{in}}$ represents the wave impedance of a spherical mode and $ka$ is on the order of $n$. For example, $d[\text{Re}(Z_{\text{in}})]/d\omega = 1.4\, d[\text{Im}(Z_{\text{in}})]/d\omega$ when $ka = n = 100$ and $Z_{\text{in}}$ equals the TM spherical mode wave impedance tuned to resonance with a series inductor. It should be emphasized that the region around $ka \approx n$ is particularly important for understanding directivity-bandwidth tradeoffs of electrically large antennas since this is the region where most of the power is radiated when directivity is maximized. However, this is precisely the region where the series RLC circuit approximation is invalid.

*B. Collin and Rothschild*

In [9], Collin and Rothschild derive the spherical mode $Q$ through a fields-based approach rather than an equivalent circuit model. Collin and Rothschild subtract the energy density associated with power flow from the total energy density to delineate 'stored' and 'radiated' components of energy. Their definitions of $Q$ were later refined by Fante [31], McLean [32], and Geyi [14]. Their analysis yields the exact same spherical mode $Q$ as the Chu circuit model, which provides supporting evidence for the validity of both approaches [10]. However, Collin and Rothschild's analysis assumes that radiated energy propagates at the speed of light, which is an unproven hypothesis [33, 34]. While this assumption is true for TEM waves in free space, it is not generally valid when there exists a field component along the direction of propagation (i.e. TE or TM modes) such as an individual spherical mode.

Applying Collin and Rothschild's logic to some other canonical problems leads to clearly unphysical results. For example, consider modes within a rectangular metallic waveguide operating above cutoff. The wave impedance is purely resistive, which provides a wideband performance (40% fractional bandwidths are typical). Rectangular waveguide modes all have a group velocity (or equivalently energy velocity here) less than the speed of light, which causes Collin and Rothschild's method to underestimate the portion of energy that is radiated and overestimate the stored energy. In fact, their analysis suggests the stored energy (non-propagating) within the waveguide above cutoff is directly proportional to the length of the waveguide. However, in practice, the bandwidth of microwave systems employing rectangular waveguides is not generally impacted by the length of the waveguides.

The same issue is found when considering radiation from the TM$_{100,m}$ spherical mode again. The field on the surface at $kr = 150$ is an interference pattern represented by the Legendre Polynomial. Over most of the surface, the field is analogous to two interfering plane waves propagating at angles $\pm 42°$ relative to the normal direction. Therefore, the radiative energy should propagate radially outwards at a velocity of 0.74c, where c equals the speed of light in free space. Of course, as the radius increases, the fields approach a TEM wave that only propagates in the radial direction. Davis et. al. [35] calculated an updated $Q = 1/(ka)^3$ for the TM$_{10}$ spherical mode that also observed radiative energy travels less than the speed of light near the antenna because power propagates at an angle relative to the normal direction. Manteghi provided a simplified derivation of Davis' $Q$ in [36]. However, Davis and Manteghi define stored energy as the difference between the total electric and magnetic energy which is problematic because this results in $Q = 0$ for any self-resonant antenna [30]. In summary, assuming the radiative component of energy propagates at the speed of light results in an overestimation of stored energy, which overestimates the radiation $Q$.

*C. Yaghjian and Best*

In [30], Yaghjian and Best consider a couple different methods of defining $Q$ of arbitrary antennas. A rigorous definition of $Q$ is defined in Eq. (80) of their paper. However, this expression simplifies to Collin and Rothschild's result when applied to the fields external to a sphere, which we argue should be updated.

Yaghjian and Best also introduce a simple approximation of the antenna $Q$ based on the frequency derivative of the impedance at resonance,

$$Q_{\text{Yaghjian}} = \frac{\omega_0 |Z_0'|}{2R_0(\omega_0)} \quad (2)$$

where $\omega_0$ corresponds to the resonant frequency where the input reactance is 0, $R_0(\omega_0)$ is the input impedance at resonance, and $|Z_0'|$ is the absolute value of the derivative of the impedance with respect to frequency. Yaghjian and Best show that this expression for $Q$ can always predict the fractional bandwidth provided the specified drop in accepted power is small enough. For example, $Q_\text{Yaghjian}$ may accurately predict the $-20$ dB bandwidth, but not the $-3$ dB bandwidth. This expression is particularly attractive because its definition is directly related to the impedance bandwidth, which is typically the end goal of calculating $Q$. Thus, the definition circumvents the challenge of dividing energy into stored and radiated components to calculate $Q$. Furthermore, $Q_\text{Yaghjian}$ is easily evaluated for arbitrary antennas, which has led to its widespread adoption.

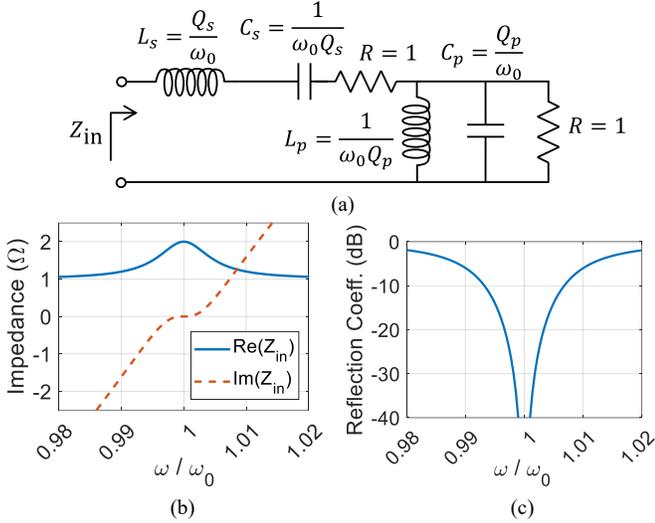

Fig. 2. (a) Example double resonant circuit model. (b) Circuit input impedance when $Q_s = Q_p = 100$. (c) Circuit reflection coefficient.

However, there is ambiguity in using (2) to calculate the $Q$ of multi-resonant antennas, as discussed in [37, 38]. For example, consider an antenna that has an input impedance given by the equivalent circuit in Fig. 2(a). The input impedance for $Q_s = Q_p = 100$ is plotted in Fig. 2(b). At resonance ($\omega = \omega_0$), taking the ratio of the stored to radiate energy suggests $Q = 100$, which is roughly twice as large as would be expected for a single resonant antenna with fractional bandwidth shown in Fig. 2(c). This agrees with the notion that there is an approximate relationship between $Q$ and bandwidth for multi-resonant designs. However, Fig. 2(b) shows that the frequency derivative at resonance is 0 which leads to $Q_\text{Yaghjian} = 0$ in (2), which is unphysical. Clearly $Q_\text{Yaghjian}$ overestimates the impedance bandwidth in this particular scenario. In fact, a matching network can always be added to an arbitrary antenna to force $Q_\text{Yaghjian} = 0$ at a particular frequency.

The fact that $Q_\text{Yaghjian}$ can fail to provide a meaningful value for multi-resonant antennas is particularly problematic for calculating directivity and bandwidth limitations of antennas. For example, [39] maximizes the directivity of an ideal spherical antenna that radiates the $TE_{10}$ and $TM_{10}$ modes using a feed network that excites the $TE_{10}$ and $TM_{10}$ modes with equal amplitude and phase. The feed automatically forces the antenna into a multi-resonant regime with $Q_\text{Yaghjian}$ near 0 even though the fractional bandwidth can be exceedingly narrow ($< 1\%$).

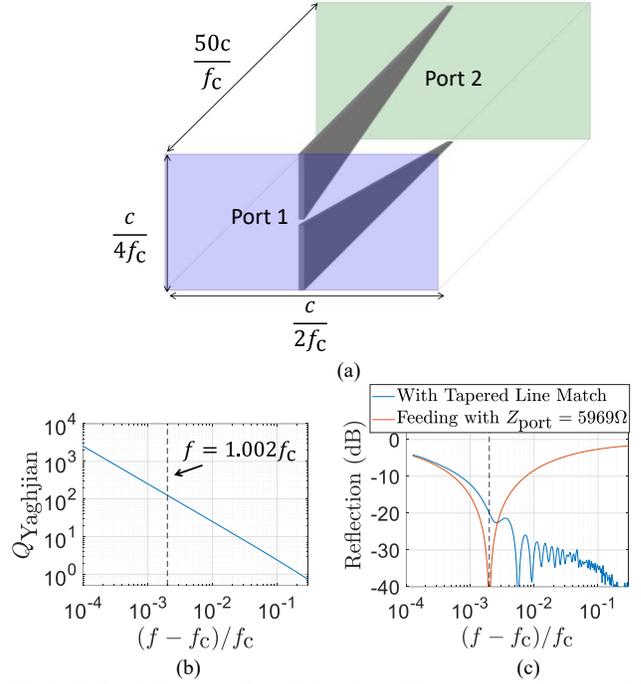

Fig. 3. (a) Dual ridge taper for wideband matching a rectangular waveguide to a 100 Ω feed. (b) $Q_\text{Yaghjian}$ of the wave impedance diverges as the operating frequency approaches cut-off. (c) Simulated reflection coefficients of the structure shown in (a) that includes a matching network (blue curve) and the scenario where the waveguide is simply fed with a port that has input impedance equal to the wave impedance at $f = 1.002 f_c$ (red curve).

Furthermore, $Q_\text{Yaghjian}$ can have limitations in predicting achievable bandwidth even when $Q_\text{Yaghjian} \gg 1$. For example, consider a simple rectangular waveguide with wave impedance $Z_\text{TE10}^\text{Rec}$, which behaves similar to the wave impedance of the spherical $TE_{100,m}$ mode. Intuitively, let us for the moment consider a fields-based definition of $Q$. We might expect the fields-based $Q$ be 0 at all frequencies above cutoff because the fields are purely propagating and there is no stored energy. However, the impedance based definition, $Q_\text{Yaghjian}$, approaches infinity as the operating frequency approaches cutoff, $f_c$, as shown in Fig. 3(b) because the wave impedance, $Z_\text{TE10}^\text{Rec} = \eta_0/\sqrt{1 - (f_c/f)^2}$, varies rapidly with frequency. A natural question to ask is whether a fields-based or impedance-based definition more accurately predicts the achievable bandwidth? If the waveguide is simply excited with a port impedance equal to the wave impedance at say $1.002 f_c$ (i.e., $Z_\text{port} = 5969$ Ω), the -20 dB bandwidth is very narrow (0.16%) which agrees with $Q_\text{Yaghjian} = 125$. However, consider instead feeding the rectangular waveguide with a lossless gradually tapered double-ridged waveguide, as shown in Fig. 3(a). The characteristic impedance at Port 1 is 100 Ω, whereas the traditional rectangular waveguide impedance at Port 2 varies rapidly with frequency near cutoff. As shown in Fig. 3(c), the reflection coefficient is less than -20 dB at all frequencies above $1.002 f_c$ (i.e., wideband behavior). In fact, a long enough double-ridged taper can provide an excellent impedance match to a rectangular waveguide at all frequencies arbitrarily close to cutoff, which is consistent with a $Q = 0$. In other words, both $Q$ definitions can be useful in this case. While $Q_{\text{Yaghjian}}$ predicts the bandwidth when there is no matching network besides a resonating inductor or capacitor, the fields-based $Q$ more accurately predicts the achievable performance with a matching network present, at least for this example. It should be emphasized that it is currently unclear how the fields-based $Q$ relates to achievable bandwidth for an arbitrary structure in general.

*D. Vandenbosch*

In [15], Vandenbosch reformulated Yaghjian and Best's rigorous definition of $Q$ in terms of integrals of the volumetric current density. Vandenbosch's analysis can also be interpreted as an extension of Geyi's definition from small antennas to electrically large structures [14]. Vandenbosch's definition, $Q_{\text{Vandenbosch}}$, is particularly attractive because it can be efficiently calculated for arbitrary geometries using the method of moments. Ref. [17] then showed that $Q_{\text{Vandenbosch}}$ can be used to bound the minimum $Q$ for an arbitrarily shaped object through convex optimization. Many studies have extended these bounds to consider various tradeoffs between size, shape, $Q$, efficiency, gain, and directivity [2, 40, 41, 42, 43, 5].

However, Vandenbosch's formula for $Q$ suffers from a few known deficiencies when $ka$ is large. As with previously mentioned definitions, it assumes that energy propagates at the speed of light, which is not generally true. Crucially, $Q_{\text{Vandenbosch}}$ is often negative for electrically large structures, which brings into question its applicability for such structures [17, 44]. A physical explanation for negative $Q_{\text{Vandenbosch}}$ values is discussed in [16]. For spherical geometries, $Q_{\text{Vandenbosch}}$ subtracts a radiative component of the energy from the energy stored on the interior of the antenna. This is unintuitive since there is no time-averaged power flow inside the sphere. Thus, $Q_{\text{Vandenbosch}}$ has an uncertainty on the order of $ka$ [16]. While this uncertainty might be insignificant for electrically small antennas, it reveals itself when analyzing electrically large structures.

For example, consider a spherical shell supporting electric currents that radiates the $TM_{10}$ spherical mode. $Q_{\text{Vandenbosch}}$ using the maximum of the electric or magnetic stored energy is analytically calculated in [16] and plotted in Fig. 4. For comparison, $Q_{\text{Hansen}}$ derived by Hansen and Collin in [12] is also plotted. Note that $Q_{\text{Hansen}}$ is identical to Collin and Rothschild's definition, but $Q_{\text{Hansen}}$ also accounts for energy stored within the spherical current shell. The two formulas ($Q_{\text{Hansen}}$ and $Q_{\text{Vandenbosch}}$) agree sufficiently well near small electrical sizes but diverge at large values of $ka$. $Q_{\text{Vandenbosch}}$ takes on increasingly negative values as $ka$ increases, whereas $Q_{\text{Hansen}}$ increases with $ka$. Ref. [16] suggests that the issue of negative $Q_{\text{Vandenbosch}}$ can be dealt with by artificially increasing $Q_{\text{Vandenbosch}}$ to 0, but this is rather arbitrary and neither physical nor valid for electrically large structures.

With respect to engineering relevancy, [17] uses $Q_{\text{Vandenbosch}}$ to calculate a bound for the maximum directivity/$Q$ ratio for antennas with arbitrary shape and size. This result is extended in [40] to also consider other tradeoffs between $Q$, directivity, radiation pattern, and placement of structures next to the antenna. Material losses are added in [2] to identify the maxim possible gain for various shapes. Ref. [5] extends [2] by decomposing the optimal currents into characteristic modes, which lends significant design/implementation insight. Setting aside the fact that $Q_{\text{Vandenbosch}}$ is unphysical for electrically large structures, [2, 40, 17, 5] have another potential limitation. These previous works limit the electric current distribution onto a specified *surface* bounding a given volume. Forcing currents to only flow on the bounding surface of a 3D object generally leads to optimal performance for electrically small structures because this maximizes the antenna size. However, forcing electric currents to only flow on the outside boundary does not always optimize performance for electrically large structures. In other words, there is no reason to believe this current distribution is optimal for antennas confined to the *volume* within the bounding surface.

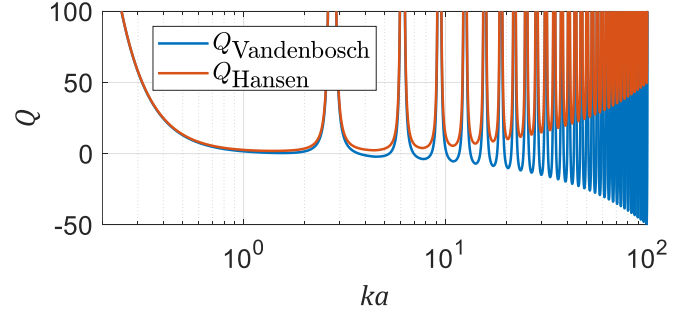

Fig. 4. Comparison of Vandenbosch's definition for $Q$ [15] and Hansen and Collins definition for $Q$ [12] of a spherical shell supporting electric current (i.e., spherical wire antenna) radiating the $TM_{10}$ mode.

For example, the infinite values of $Q$ in Fig. 4 correspond to resonances where the inward directed wave impedance is 0 Ω. If the goal is to maximize bandwidth of the $TM_{10}$ mode, it is better to use a $ka = 1.5$ sized sphere compared to say $ka = 100$. In other words, the bounds reported in [2, 40, 17, 5] should be interpreted as the optimal performance for a class of antennas that only support electric currents on the specified surfaces. This should not be confused with bounding the performance of arbitrary antennas confined to the volume within the surface.

*E. Comparison of Q Definitions*

Table 1 highlights the differences between the various $Q$ definitions by comparing their values for the $TM_{100,m}$ mode at different values of $ka$. $Q_{\text{Yaghjian}}^{\text{L Match}}$ corresponds to (2) when a series inductor is used to match the $TM_{100,m}$ wave impedance to a resistive load while $Q_{\text{Yaghjian}}^{\text{Match Net.}}$ uses a matching network between the load and $TM_{100,m}$ wave impedance to minimize its $Q$. Note that $Q_{\text{Vandenbosch}}$ corresponds to a spherical shell of electric currents that supports internal stored energy, which is a different scenario than $Q_{\text{Chu}}$. $Q_{\text{Pfeiffer}}$ denotes the new $Q$ definition that will be introduced in Section IV. Intuitively, we expect a high $Q$ when $ka < 100$ and a low, but positive $Q$ when $ka > 100$. Only $Q_{\text{Pfeiffer}}$ satisfies these crucial criteria at all tabulated values of $ka$.

TABLE I
COMPARISON OF DIFFERENT $Q$ DEFINITIONS WHEN APPLIED TO THE $TM_{100,m}$ SPHERICAL MODE

| $ka$ | 95 | 101 | 110 |
|---|---|---|---|
| $Q_{\text{Chu}}$ [8] | 173 | 87 | 65 |
| $Q_{\text{Collin}}$ [9] | 173 | 87 | 65 |
| $Q_{\text{Yaghjian}}^{\text{L Match}}$ [30] | 90 | 10 | 2.4 |
| $Q_{\text{Yaghjian}}^{\text{Match Net.}}$ [30] | 0 | 0 | 0 |
| $Q_{\text{Vandenbosch}}$ [15] | 118 | 1.7 | -20 |
| **$Q_{\text{Pfeiffer}}$** | **98** | **1.3** | **0.01** |

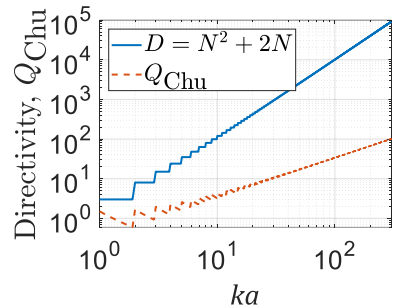

Fig. 5. Directivity and $Q$ of the spherical mode excitation as calculated by Harrington when the Chu circuit model is used to calculate $Q_n$ of the $n$th order mode.

## III. Issues with Previous Directivity-$Q$ Tradeoffs

Next, we quickly review issues with previous analyses that investigate tradeoffs between antenna directivity and bandwidth.

### A. Harrington

In [25], Harrington calculates the maximum directivity of an antenna that supports $N$ modes by deriving the modal coefficients that maximize the directivity. He also calculates the corresponding antenna $Q$, which is a function of the quality factor of each spherical mode derived using the Chu circuit model. Harrington states that when $N < ka$, the radiation quality factor is on the order of unity and the directivity equals $N^2 + 2N$ (i.e., approximately 100% aperture efficiency). This suggests it is possible to realize approximately 100% aperture efficiency antennas with broad bandwidth. However, Harrington approximated the spherical mode quality factor using Chu's approximate series RLC circuit, which is not generally accurate for all spherical modes. If the 'rigorous' version of spherical mode $Q_{\text{Chu}}$ from the Chu circuit model is used instead, the quality factor grows without bound as $ka$ increases. For example, Fig. 5 plots the directivity and $Q_{\text{Chu}}$ when Harrington's optimal combination of spherical modes are excited. Fig. 5 suggests that high directivity antennas with approximately 100% aperture efficiency must be extremely narrowband, which defies conventional wisdom.

Undoubtedly, the definition of $Q$ needs updating for Harrington's analysis to provide a correct understanding of the tradeoffs between antenna directivity and bandwidth. But beyond this, Harrington's restriction on the number of spherical modes employed for radiation is arbitrary and leads to a poorly defined limit. Constraining the number of spherical modes to be $N \leq ka$ creates a staircase expression for the $(D = N^2 + 2N)$ and $Q$ that abruptly jumps in a non-physical wave whenever $ka$ is an integer. Furthermore, Harrington calculates that the 'optimal' power distribution across all spherical modes increases with mode order, $n$, as $2n + 1$ and then abruptly reduces to 0 when $n > ka$, which will be shown to be unnecessary. Surely, a smoother relationship between radiated power and spherical mode index could result in both larger directivity and bandwidth. This insight has led to heuristic definitions of the maximum practical directivity [45, 46]. In fact, calculating a rigorous limit on the maximum achievable directivity for a specified bandwidth was only recently solved in [28, 29] for the general case when the antenna is not electrically small. However, [28, 29] should be updated to use the new definition of $Q$ that is proposed here to ensure physically meaningful results when $ka \gg 1$.

### B. Fante and Geyi

In [8], Chu calculates the spherical modal coefficients that maximize the ratio $D/Q$ (i.e., directivity-bandwidth product) for omni-directional antennas that have azimuthally symmetric radiation in the $\theta = 90°$ plane. In [26], Fante extends this analysis to consider pencil-beam antennas. In [27], Geyi points out an error in Fante's analysis and derives a simple updated expression for the spherical modal coefficients that maximize the $D/Q$ ratio. Today, the $D/Q$ ratio remains a commonly employed metric for characterizing antennas [2, 40].

However, maximizing $D/Q$ tends to yield a spherical mode excitation with $Q \ll 1$ for antennas that are not electrically small. Using this $D/Q$ limit to estimate the maximum achievable directivity for antennas with moderate sizes (e.g., $ka > 2$) and $Q$ (e.g., $Q > 1$) results in a loose bound in the sense that it significantly overestimates the achievable directivity. Furthermore, previous results all rely on Chu's definition for $Q$, which should be updated. It will be shown the $D/Q$ limit becomes even looser when the spherical mode $Q$ is updated using our updated definition.

## IV. New $Q$ Factor Definition for Spherical Modes

In this section, a new definition of antenna $Q$ based on spherical mode radiation is postulated. The delineation between stored and radiated energy follows naturally from an exact transmission line model that represents the fields.

### A. Equivalent Transmission Line Model

Consider an arbitrary antenna with sources contained within the spherical region $r < a$. The field external to $r = a$ can be written as a superposition of $TE_{nm}$ and $TM_{nm}$ spherical modes where $n$ is the order of the spherical Bessel function and $m$ is the azimuthal variation [47]. The radiated and stored energy is the sum of the contributions from each mode since all spherical modes are orthogonal. Furthermore, TE radiation is simply the dual of TM. Therefore, it is sufficient to analyze the stored and radiated energy in the TM$_{nm}$ modes with the understanding that the analysis can be extended to all other modes through duality and/or symmetry.

Outward propagating TM$_{nm}$ waves have fields given by,

$$E_\theta = C_{nm} \frac{j\eta_0}{rk} \frac{dP_n^m(\cos\theta)}{d\theta} \frac{d[krh_n^{(2)}(kr)]}{dr} g(m\phi)$$

$$E_\phi = C_{nm} \frac{j\eta_0}{rk\sin(\theta)} P_n^m(\cos\theta) \frac{d[krh_n^{(2)}(kr)]}{dr} \frac{dg(m\phi)}{d\phi}$$

$$E_r = C_{nm} \frac{j\eta_0}{k} \frac{n(n+1)}{r^2} P_n^m(\cos\theta)[krh_n^{(2)}(kr)]g(m\phi) \quad (3)$$

$$H_\theta = -C_{nm} \frac{P_n^m(\cos\theta)}{r\sin(\theta)} [krh_n^{(2)}(kr)] \frac{dg(m\phi)}{d\phi}$$

$$H_\phi = C_{nm} \frac{1}{r} \frac{dP_n^m(\cos\theta)}{d\theta} [krh_n^{(2)}(kr)] g(m\phi)$$

where $\eta_0 = \sqrt{\mu_0/\varepsilon_0}$ is the free space wave impedance, $g(m\phi)$ is either $\cos(m\phi)$ or $\sin(m\phi)$, $P_n^m$ is the associated Legendre polynomial, $h_n^{(2)}$ is the spherical Hankel function of the second kind, $C_{nm}$ is an arbitrary constant with units of ampere, and an $e^{j\omega t}$ time convention is used.

It is known that these spherical modes can be viewed as propagating within a waveguide [47], which in turn has a model based on transmission line theory. The wave impedance of a mode ($E_\theta/H_\phi$) corresponds to the outward looking impedance on the transmission line, which in general is a function of $r$, but not $\theta$ or $\phi$. However, the transmission line's propagation constant, $\beta(r)$, and characteristic impedance, $Z_0(r)$, are yet to be determined. The propagation constant and characteristic impedance are calculated by first defining a complex voltage, $V_{nm}$, and complex current, $I_{nm}$, that are proportional to the electric and magnetic fields, respectively. As discussed in [48, 49], while there is some arbitrariness in normalizing the voltage and current in terms of the waveguide fields, the important principle is to ensure there is consistency between the complex power and impedance in the transmission line model and in the waveguide. The voltage and current are normalized here as [48],

$$V_{nm} I_{nm}^* = \int_{4\pi} \bar{E} \times \bar{H}^* \cdot \hat{r}\, r^2\, d\Omega$$
$$\left(\frac{V_{nm}}{I_{nm}}\right) \bar{H} = \hat{r} \times \bar{E} \quad (4)$$

where the limits of integration are over the unit sphere, $\hat{r}$ is the radially directed unit vector, and $*$ denotes complex conjugate. Thus, it can be verified by inspection that (4) is satisfied when the voltage and current are given by,

$$V_{nm} = jC_{nm}\alpha_{nm} \frac{\eta_0}{k} \frac{d[krh_n^{(2)}(kr)]}{dr}$$
$$I_{nm} = C_{nm}\alpha_{nm}[krh_n^{(2)}(kr)] \quad (5)$$

where $\alpha_{nm}$ is a normalization factor that enforces power carried by the equivalent transmission line voltage and current equals power flowing through the waveguide cross section (i.e., spherical shell with constant $r$),

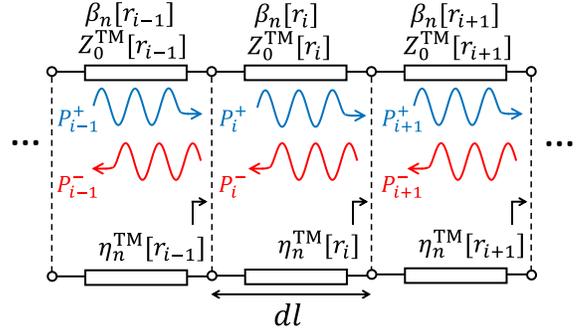

Fig. 6. Discretized transmission line model for calculating spherical mode $Q$. $Z_0^{TM}$ corresponds to the characteristic impedance of the transmission line section while $\eta_n^{TM}$ is the wave impedance.

$$\alpha_{nm} = \sqrt{\int_{4\pi} \left(\frac{dP_n^m(\cos\theta)g}{d\theta}\right)^2 + \frac{\left(\frac{dP_n^m(\cos\theta)g}{d\phi}\right)^2}{\sin(\theta)^2}\, d\Omega}$$

$$= \begin{cases} \sqrt{\dfrac{4\pi n(n+1)}{2n+1}} & m = 0 \\[2mm] \sqrt{\dfrac{2\pi n(n+1)}{2n+1}\dfrac{(n+m)!}{(n-m)!}} & m \neq 0 \end{cases} \quad (6)$$

Differentiating $I_{nm}$ and $V_{nm}$ with respect to $r$ leads to the following coupled differential equations,

$$\frac{dI_{nm}}{dr} = -\frac{jk}{\eta_0} V_{nm}$$
$$\frac{dV_{nm}}{dr} = -\frac{j\eta_0}{k}\left(k^2 - \frac{n(n+1)}{r^2}\right) I_{nm} \quad (7)$$

where the second derivative of the spherical Hankel function is replaced with itself using the definition of the Riccati-Bessel functions [50],

$$\frac{d^2[krh_n^{(2)}(kr)]}{dr^2} + \left(k^2 - \frac{n(n+1)}{r^2}\right)[krh_n^{(2)}(kr)] = 0 \quad (8)$$

Comparing (7) with the conventional telegrapher's equations of a lossless transmission line,

$$\frac{dI}{dr} = -j\omega C V$$
$$\frac{dV}{dr} = -j\omega L I \quad (9)$$

allows for defining a radially varying waveguide with propagation constant and characteristic impedance given by,

$$\beta_n(r) = k\sqrt{1 - \frac{n(n+1)}{(kr)^2}} = \omega\sqrt{LC} \quad (10)$$

$$Z_0^{TM}(r) = \frac{\eta_0 \beta_n(r)}{k} = \sqrt{L/C} \quad (11)$$

Thus, the fields of spherical mode propagation can be exactly modelled using a non-uniform 1D waveguide with voltage and current that propagate in the forward and backward directions as $\exp(\pm j\beta_n r)$. Note that the propagation constant and characteristic impedance of the transmission line are independent of the azimuthal variation, $m$. In the following, we use traditional transmission line theory to relate the characteristic impedance and total fields ($\bar{E}, \bar{H}$) to forward (+) and backward (−) propagating field components ($\bar{E}^{+,-}, \bar{H}^{+,-}$).

Strictly speaking, the fields in a waveguide only propagate as $\exp(\pm j\beta_n r)$ when the waveguide does not vary with position. However, this does not restrict the analysis. Traditional mode matching techniques commonly solve for the transverse fields of non-uniform waveguides by discretizing the waveguide into short uniform sections [47, 51], each of which has differential length equal to $dl$, as shown in Fig. 6. Unambiguously, once the transverse fields are solved, the radial components can be readily evaluated by taking the curl of the transverse fields. The propagation constant and impedance of each waveguide section are given by (10) and (11), where $r$ is evaluated at the midpoint of the waveguide section. In the limit that $dl \to 0$, the fields in the discretized waveguide are exactly equal to those in spherical mode radiation. In other words, our transmission line model can be thought of as an infinitely precise mode matching analysis since spherical mode radiation is analytic.

The transmission line model provides insight into how a spherical mode propagates. Like propagation in a conventional waveguide, there is a transverse wavenumber that satisfies the boundary conditions and determines the propagation constant through the dispersion relation. Comparing the propagation constant in (10) with that of a conventional waveguide suggests the term, $\sqrt{n(n+1)}/r$, should be interpreted as the transverse wavenumber at the radius, $r$. Previously, spherical modes were thought to have an ambiguous cutoff frequency near $kr \approx n$ because this is roughly where the wave impedance transitions from being primarily reactive to resistive [47]. In contrast, the equivalent transmission line model introduced here provides a precise cutoff frequency when the transverse wavenumber equals the free space wavenumber, $k = \sqrt{n(n+1)}/r$. It might seem counterintuitive that power can flow (not necessarily propagate) through a waveguide section below cutoff. However, waveguide filters commonly tunnel energy through finite sections operating below cutoff to reach a propagating region.

Conceptually, our analysis suggests propagation of a spherical mode is analogous to propagation through a tapered rectangular waveguide. The outward looking wave impedance is in general different than the characteristic impedance at any given location. Therefore, forward and backward directed waves exist at every position [47]. In the region where $kr < \sqrt{n(n+1)}$ the analogous tapered waveguide is below cutoff with a superposition of evanescently decaying/growing fields. As we move outwards, the waveguide walls widen and the mode propagates above cutoff when $kr > \sqrt{n(n+1)}$. As we continue to move outwards, the waves bend toward the normal direction and the reflected/inward propagating wave amplitude decreases. In the limit that $kr \to \infty$, the mode transitions to a TEM mode outwardly propagating in the radial direction. Note that the tapered rectangular waveguide analogy is only conceptual and there is not a rigorous one-to-one relationship between spherical modes and a tapered waveguide. A rectangular waveguide taper couples mode together, whereas spherical modes all propagate outwards independently of one another [51].

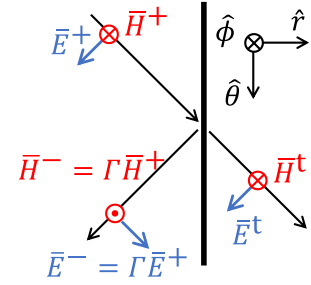

Fig. 7. Decomposition of a TM$_{n0}$ polarized propagating wave into incident ($\bar{E}^+, \bar{H}^+$), reflected ($\bar{E}^-, \bar{H}^-$), and transmitted ($\bar{E}^t, \bar{H}^t$) components.

This transmission line model also provides a straightforward interpretation of stored ($U_s$) and radiated ($U_r$) energy density,
$$U = U_s + U_r \tag{12}$$
where the total energy density, $U$, can also be decomposed into electric ($U^e$) and magnetic ($U^m$) energy components,
$$U = U^e + U^m = \frac{\varepsilon_0 |\bar{E}|^2}{4} + \frac{\mu_0 |\bar{H}|^2}{4} \tag{13}$$
The total energy ($W$) is simply the integral of the energy density ($U$) over all space. We define the radiated energy at any given section to be the component of the energy that carries power outward through a propagating wave (above cutoff). The stored energy is everything else.

When the differential transmission line segment is below cutoff, the fields exponentially decay/grow with position and there is no radial 'propagation'. Thus, there is no radiative or propagating component of the energy such that stored energy equals total energy.

A key consequence of modelling a spherical wave using a transmission line is the ability to consider the effect of cutoff and reflections at each differential section of the transmission line in a novel but intuitive way. Above cutoff, the fields are composed of known forward and backward propagating TM polarized waves as depicted in Fig. 7. As with every waveguide above cut-off, the ratio of the tangential electric field of the backward to forward propagating wave at any point on the transmission line can be represented by a reflection coefficient $\Gamma(r)$ that is a function of the transmission line characteristic impedance and the input impedance. We therefore define $\Gamma(r)$ as,
$$\Gamma(r) = \frac{\eta_n^{\text{TM}}(r) - \text{Re}\left(Z_0^{\text{TM}}(r)\right)}{\eta_n^{\text{TM}}(r) + \text{Re}\left(Z_0^{\text{TM}}(r)\right)} \tag{14}$$
where Re denotes real part of the argument,
$$\eta_n^{\text{TM}}(r) = \frac{j\eta_0}{krh_n^{(2)}(kr)} \frac{d\left[rh_n^{(2)}(kr)\right]}{dr} \tag{15}$$
is the wave impedance of the outward propagating spherical TM mode, and $Z_0^{\text{TM}}(r)$ is the characteristic impedance defined in (11). The reflection coefficient in (14) is defined using the real part of the transmission line impedance, which is unconventional. The reflection coefficient is typically only defined in the context of propagating fields above cutoff where the characteristic impedance, $Z_0^{\text{TM}}(r)$, is purely real. Therefore, taking the real part of $Z_0^{TM}(r)$ does not affect the current discussion because $Z_0^{TM}(r)$ is purely real above cutoff anyway. Below cutoff, $Z_0^{TM}(r)$ is purely imaginary and the reflection coefficient in (14) is unity. It will be shown that this definition for the reflection coefficient forces the stored energy to be equal to the total energy as we discussed previously. In other words, the definition for the reflection coefficient in (14) simplifies subsequent expressions for stored and radiated energy by allowing them to be valid below and above cutoff.

Referencing Fig. 7 above cutoff, the incident and reflected voltages ($V_0^{+,-}(r)$) and currents ($I_0^{+,-}(r)$) on the transmission line have the conventional relationship with the total voltage and current,

$$V = V_0^+ + V_0^- = V_0^+(1+\Gamma)$$
$$I = I_0^+ - I_0^- = I_0^+(1-\Gamma) = V_0^+(1-\Gamma)/Z_0^{TM} \quad (16)$$

Analogous definitions exist for the incident and reflected components of the tangential electric ($E_{\theta,\phi}^{+,-}$) and magnetic ($H_{\theta,\phi}^{+,-}$) fields. In other words, $E_{\theta,\phi} = E_{\theta,\phi}^+(1+\Gamma)$ and $H_{\theta,\phi} = H_{\theta,\phi}^+(1-\Gamma)$. Furthermore, the incident and reflected components of the radially directed field ($E_r^{+,-}$) are also uniquely defined by referencing Fig. 7,

$$E_r = E_r^+ - E_r^- = E_r^+(1-\Gamma) \quad (17)$$

where $E_r$ is given by (3).

It is important to note that forward propagating waves on all conventional waveguides above cutoff have equal electric and magnetic energy components when the corresponding energy densities are integrated over the waveguide cross section. Appendix B demonstrates that the same is true here for outward or inward propagating waves on the effective spherical mode waveguide,

$$\frac{\mu_0}{4}\int_{4\pi}|\bar{H}^+|^2 r^2 d\Omega = \frac{\varepsilon_0}{4}\int_{4\pi}|\bar{E}^+|^2 r^2 d\Omega$$
$$= \frac{\mu_0}{4}\int_{4\pi}\left(\left|\frac{H_\phi}{1-\Gamma}\right|^2 + \left|\frac{H_\theta}{1-\Gamma}\right|^2\right)r^2 d\Omega \quad (18)$$
$$= \frac{\varepsilon_0}{4}\int_{4\pi}\left(\left|\frac{E_r}{1-\Gamma}\right|^2 + \left|\frac{E_\theta}{1+\Gamma}\right|^2 + \left|\frac{E_\phi}{1+\Gamma}\right|^2\right)r^2 d\Omega$$

It is the interaction between forward propagating and reflected waves that results in the TM mode storing more electric energy than magnetic energy. An analogous example is the energy on a mismatched transmission line oscillates between predominately electric and magnetic as the observation plane moves along the line. Eqn. (18) being true provides some validation that both the transverse and radial components of the field are accurately captured in the derived transmission line model.

Next, consider the diagram in Fig. 8 representing the superposition of forward and backward propagating waves in the waveguide above cutoff. The net time-averaged power flow ($P(r)$) at all $r$ is related to powers of outward ($P^+(r)$) and inward ($P^-(r)$) propagating waves as,

$$P(r) = P^+(r) - P^-(r) = P^+(r)(1-|\Gamma(r)|^2)$$
$$= \frac{1}{2}\int_{4\pi}\text{Re}(\bar{E}\times\bar{H}^*)\cdot\hat{r} r^2 d\Omega \quad (19)$$

where

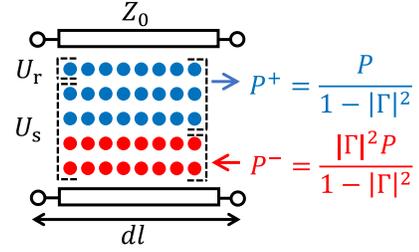

Fig. 8. Diagram representing energy propagating in the forward (blue circles) and backward (red circles) directions along a waveguide section above cutoff. The stored energy density is defined as the stationary portion of energy that does not carry power. The radiative energy is the remainder that carries power in the forward direction. This is analogous to a resonant transmission line section.

$$P^{+,-}(r) = \frac{1}{2}\int_{4\pi}\text{Re}(\bar{E}^{+,-}\times\bar{H}^{*+,-}\cdot\hat{r}) r^2 d\Omega \quad (20)$$

Each blue and red circle in Fig. 8 corresponds to a unit of energy flowing in the forward (i.e., to the right) and backward (i.e., to the left) directions, respectively. Despite the spatial variation, power is conserved. The blue circles are divided into two segments: (a) those that cancel red circles and (b) the remaining circles that carry energy forward. Since the lower segment of blue and red circles do not carry any net energy in the forward or backward directions, we denote these as 'stored' energy, akin to the physics of a standing wave. The upper section of blue circles that carry net energy forward are thus 'radiative' energy. More precisely, the radiative energy density within a differential spherical shell at $r$ is written as,

$$\int_{4\pi} U_r(r) r^2 d\Omega = \left(\frac{P^+(r)-P^-(r)}{P^+(r)+P^-(r)}\right)\int_{4\pi} U(r) r^2 d\Omega$$
$$= \left(\frac{1-|\Gamma(r)|^2}{1+|\Gamma(r)|^2}\right)\int_{4\pi} U(r) r^2 d\Omega \quad (21)$$

and the stored energy density within the shell is given by,

$$\int_{4\pi} U_s(r) r^2 d\Omega = \left(\frac{2P^-(r)}{P^+(r)+P^-(r)}\right)\int_{4\pi} U(r) r^2 d\Omega$$
$$= \left(\frac{2|\Gamma(r)|^2}{1+|\Gamma(r)|^2}\right)\int_{4\pi} U(r) r^2 d\Omega \quad (22)$$

Note that (21) and (22) are also valid below cutoff since all energy is stored due to $\Gamma(r) = 1$. When the waveguide is impedance matched such that $P^- = 0$, all of the energy is radiative. When, $P^- = P^+$ there is no energy propagation, and all energy is stored. A diagram summarizing the different regions of interest for analyzing spherical mode radiation is shown in Fig. 9.

A natural interpretation follows. Energy in each waveguide section propagates in the forward and backward directions with equal speeds but opposite directions. The speed of this propagating energy at each position is equal to the total power directed toward that region divided by the total energy density in the same region,

$$v_r(r) = \frac{P^+(r)+P^-(r)}{\int_{4\pi} U(r) r^2 d\Omega} = \frac{P}{\int_{4\pi} U_r(r) r^2 d\Omega} \quad (23)$$

We define this velocity, $v_r$, as the 'radiative energy velocity'. An intuitive discussion on how this new radiative energy velocity definition is related to more commonplace group and energy velocities is provided in Appendix A. Since the net power is carried only by the component of energy that is radiative, the net power can also be expressed in terms of the stored energy as,

$$P = v_r(r) \int_{4\pi} [U(r) - U_s(r)] \, r^2 d\Omega \quad (24)$$

Rearranging (24), the energy density stored within a differential spherical shell at $r$ can be represented as,

$$\int_{4\pi} U_s(r) \, r^2 d\Omega = \int_{4\pi} U(r) \, r^2 d\Omega - P/v_r(r) \quad (25)$$

This expression has the same form as that given by Collin and Rothschild [9], but with the speed of light replaced by the more representative radiative energy velocity.

The present analysis also offers insight into discussion related to the reactive near field region, radiative near field region (i.e., Fresnel region) and far-field region surrounding an antenna. For a given spherical mode, $n$, the fields are below cutoff when $kr < \sqrt{n(n+1)}$. Thus, we could also interpret this region as the reactive near field since energy doesn't propagate, but rather it tunnels outwards. However, the definition for the reactive near field remains ambiguous for a general antenna that simultaneously excites multiple spherical modes, since some modes are below cutoff and others are above cutoff. The transmission line model also provides insight into how the far-field components of the field should be interpreted. Appendix A demonstrates that electromagnetic energy generally propagates slower than the speed of light near the antenna where the fields are not TEM. The energy velocity asymptotically approaches the speed of light as $kr$ goes to infinity. When considering radiation from a Hertzian dipole, the term that decays as $1/r$ is typically associated with the radiative far field [35]. However, this interpretation is overly simplistic since the increasing energy velocity with radius results in the outward propagating portion of the field to decay faster than $1/r$ even in the region well above cutoff where a very small portion of the energy is stored.

Thus far, only the total stored energy (i.e., sum of the electric and magnetic energies) has been considered. However, the same logic is used to define stored electric and magnetic energy by simply replacing $U$ with $U^{e,m}$ in (22). This fact will be used to define the $Q$ of the non-resonant TE or TM modes in the following subsection. There is no assumption that the radiated electric and magnetic energy densities are equal for all $r$ (both above and below cutoff), in contrast to Collin and Rothschild's analysis [9].

### B. Spherical Mode Q Definition

Now that stored and radiated energy are defined, the quality factors for various spherical modes can be evaluated. The $Q$ is typically defined as,

$$Q = \frac{2\omega \max(W_s^e, W_s^m)}{P} \quad (26)$$

where the superscripts $e$ and $m$ denote electric and magnetic components of the energy. The stored energy ($W_s^{e,m}$) is defined as,

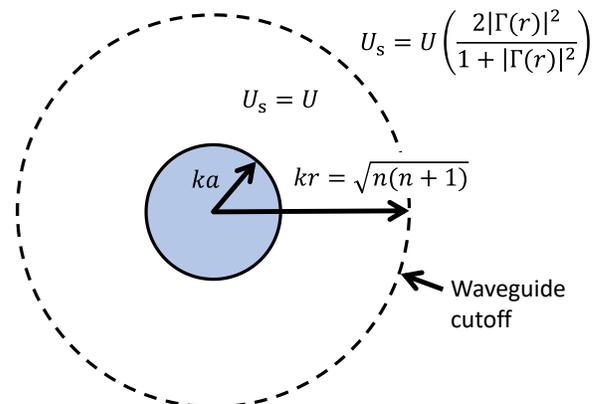

Fig. 9. The different regions of interest for spherical mode radiation.

$$W_s^{e,m} = \int_a^\infty \int_{4\pi} U_s^{e,m} \, r^2 d\Omega \, dr \quad (27)$$

It is straightforward to show that the stored electric energy density within every differential spherical shell is greater than the stored magnetic energy density for TM radiation, which implies that $W_s^e > W_s^m$ for each mode. To derive $Q$, the fields from (3) are inserted into (22). The result is inserted into (27) and (26) to provide the quality factor of the $TM_{nm}$ mode,

$$Q_n^{TM} = Q_n^{TE} = \int_{ka}^\infty \left( n(n+1) \left| h_n^{(2)}(\rho) \right|^2 \right.$$
$$\left. + \left| \frac{d[\rho h_n^{(2)}(\rho)]}{d\rho} \right|^2 \right) \left( \frac{2|\Gamma(\rho)|^2}{1 + |\Gamma(\rho)|^2} \right) d\rho \quad (28)$$

where $\rho = kr$, and $\Gamma(\rho)$ is given by (14), which in turn depends on $\eta_n^{TM}(r)$ from (15) and $Z_0^{TM}(r)$ from (11). TE spherical modes have an identical $Q_n$ as TM due to duality.

The $n = 1$ mode is of particular interest for electrically small antennas. The integral in (28) can be evaluated by replacing $h_1^{(2)}(\rho)$ with $e^{-j\rho}(j - \rho)/\rho^2$ and carrying out the integration. The integral simplifies when $ka < \sqrt{2}$ (below cutoff),

$$Q_{n=1\{ka<\sqrt{2}\}}^{TM} = Q_{n=1\{ka<\sqrt{2}\}}^{TE} = \frac{1}{(ka)^3} + \frac{1}{ka} - ka$$
$$+ \frac{\pi}{4} \left( \sqrt{99 + 47\sqrt{3}} + \sqrt{99 - 47\sqrt{3}} - 12\sqrt{2} \right)$$
$$\approx \frac{1}{(ka)^3} + \frac{1}{ka} - ka + 0.51 \quad (29)$$

This expression for $Q$ agrees with the definition from Chu to the first two leading orders, which suggests that the well-known Chu limit is valid when $ka \ll 1$. However, when $ka$ approaches unity, there is a notable difference. For example, Chu's definition for $Q$ is 32% larger than the definition in (29) when $ka = 1$.

Next, the $Q$ is analyzed when equal powers are radiated in the $TE_{nm}$ and $TM_{nm}$ modes such that $W_s^e = W_s^m$. This is another important scenario because this excitation provides the minimum $Q$ and maximum directivity [25]. From duality, the stored electric energy when TE and TM modes are radiated with the same power equals the sum of the electric and magnetic energy of only the TM radiation. Integrating this energy density over the unit sphere provides the following definition of $Q_n^{\text{TE+TM}}$,

$$Q_n^{\text{TE+TM}} = \int_{ka}^{\infty} \left[ \left(1 + \frac{n(n+1)}{\rho^2}\right) |\rho h_n^{(2)}(\rho)|^2 \right. \\ \left. + \left|\frac{d[\rho h_n^{(2)}(\rho)]}{d\rho}\right|^2 \right] \left(\frac{|\Gamma(\rho)|^2}{1+|\Gamma(\rho)|^2}\right) d\rho \quad (30)$$

where the superscript TE+TM denotes the case with equal power radiated by the TE and TM modes. Again, consider the lowest order mode, $n=1$,

$$Q_{n=1}^{\text{TE+TM}}\big|_{\{ka<\sqrt{2}\}} = \frac{1}{2(ka)^3} + \frac{1}{ka} - ka \\ + \frac{\pi}{2}\left(\sqrt{6+2\sqrt{3}} + \sqrt{6-2\sqrt{3}} - 3\sqrt{2}\right) \\ \approx \frac{1}{2(ka)^3} + \frac{1}{ka} - ka + 0.67 \quad (31)$$

The definitions of $Q$ in (28) and (30) are similar to the result from Collin and Rothschild [9], with the primary difference being the energy density is modified by the $\Gamma(\rho)$ term. Unfortunately, a simple closed form solution to the integrals in (28) and (30) was not found, so they are computed numerically. The numerical integration is trivial though since the stored energy density rapidly converges to 0 when $\rho > \sqrt{n(n+1)}$ (above cutoff). Note that the integrand has a discontinuity in the first derivative at cut-off, so we break the integral over $\rho$ into two segments when $ka < \sqrt{n(n+1)}$: $[ka, \sqrt{n(n+1)}]$ and $[\sqrt{n(n+1)}, \infty]$ to improve numerical convergence.

Fig. 10 and Fig. 11 plot some of the spherical mode quality factors as a function of $ka$. The $Q$ based on the transmission line model proposed here is denoted $Q_{\text{Pfeiffer}}$. The curves have a clear knee when the antenna size is larger than the cutoff frequency, $ka > \sqrt{n(n+1)}$. The quality factor using the circuit model defined by Chu ($Q_{\text{Chu}}$) and the impedance derivative defined by (2) ($Q_{\text{Yaghjian}}$) are also plotted for reference. The input impedance for $Q_{\text{Yaghjian}}^{\text{TE+TM}}$ assumes a hypothetical antenna excites the TE and TM mode wave impedances in series. An ideal transformer at the input of the TE wave impedance ensures equal power is radiated by the TE and TM modes. All three $Q$ definitions agree when $ka \ll n$. However, $Q_{\text{Chu}}$ derived from the Chu circuit model is much larger than $Q_{\text{Yaghjian}}$ and $Q_{\text{Pfeiffer}}$ when $ka$ is on the order of $n$ or larger, as expected. The fact that the $Q$ based on the frequency derivative of the input impedance ($Q_{\text{Yaghjian}}$) is generally in close agreement with the $Q$ based on stored energy ($Q_{\text{Pfeiffer}}$) for moderate to large values of $Q$ is a satisfying result since it suggests the field-based $Q_{\text{Pfeiffer}}$ is closely related to the impedance bandwidth.

There is disagreement between $Q_{\text{Pfeiffer}}$ and $Q_{\text{Yaghjian}}$ near cutoff when $ka$ and $n$ increase. For example, when only the TE$_{n=100}$ mode is radiated and $ka=101$, $Q_{\text{Pfeiffer}}^{\text{TE}} = 1.3$ while $Q_{\text{Yaghjian}}^{\text{TE}} = 10.0$. This discrepancy between $Q_{\text{Pfeiffer}}^{\text{TE}}$ and $Q_{\text{Yaghjian}}^{\text{TE}}$ can be understood by referring to the rectangular waveguide analogy previously discussed in Section II C. The wave impedance of the TE$_{n=100}$ mode closely resembles the wave impedance of a rectangular waveguide. Slightly above cutoff, the wave impedance rapidly changes with frequency. Therefore, there is a relatively narrow bandwidth if the wave impedance is matched to a resistive load using a single inductor or capacitor, which is in agreement with $Q_{\text{Yaghjian}}^{\text{TE}} = 10.0$. However, if we allow ourselves to use a more complicated matching network using a combination of a parallel LC resonator and the dual ridge tapered waveguide from Fig. 3(a), it is possible to impedance match the TE$_{n=100}$ mode to $100\,\Omega$ with better than 20 dB mismatch loss at all frequencies above $ka = 101$ (details are omitted for brevity). This very wideband matching behavior is consistent with the low $Q^{\text{TE}}_{\text{Pfeiffer}} = 1.3$. Thus, $Q_{\text{Pfeiffer}}$ seems to provide better insight into the achievable bandwidth with an elaborate matching network, but additional research is needed to evaluate whether this is a general result or a special case. In contrast, $Q_{\text{Yaghjian}}$ is expected to better accurately represent the bandwidth for scenarios where there exists a simple resonating inductor/capacitor and a single resonance [37].

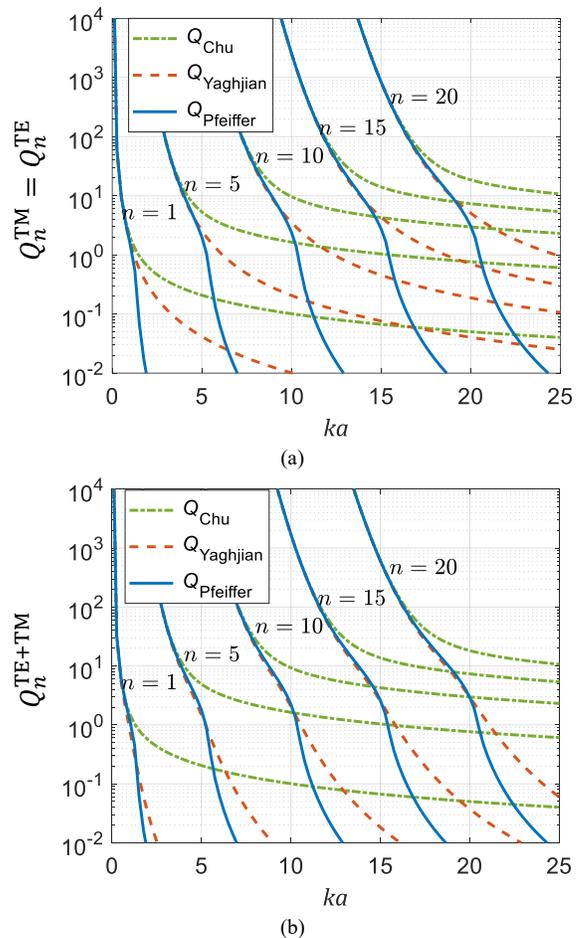

Fig. 10. Different definitions of $Q_n$ for radiation from the non-resonant TM (or TE) modes in isolation (a) and the resonant combination of TE$_{nm}$ and TM$_{nm}$ modes (b).

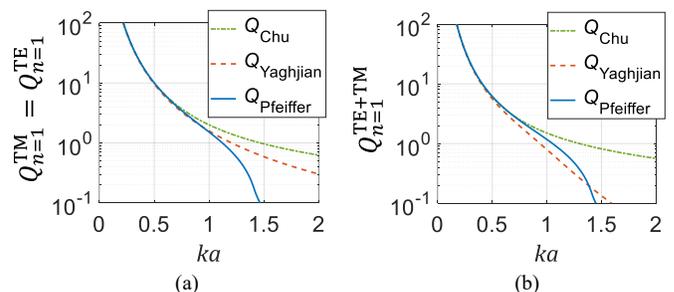

Fig. 11. Different definitions of $Q_{n=1}$ of the lowest order mode for radiation from the non-resonant TM (or TE) modes in isolation (a) and the resonant combination of TE$_{1m}$ and TM$_{1m}$ modes (b).

## V. Optimal Tradeoff Between Directivity and $Q$

Attention is now turned to find the maximum directivity of an arbitrary antenna that has a fixed size and bandwidth. This is an immediate application of applying the updated spherical mode $Q$'s.

As previously discussed, Harrington's analysis arbitrarily truncates the number of spherical harmonics, which results in a suboptimal 'bound' relating antenna size to directivity and bandwidth [25]. Geyi addresses this deficiency by showing the maximum $D/Q$ ratio equals [27],

$$\max \frac{D}{Q} = \sum_{n=1}^{\infty} \frac{2n+1}{Q_n^{\text{TE+TM}}} \tag{32}$$

This correct expression allows for the participation of modes that were previously truncated, but it nonetheless is dependent on the definition of $Q$. When Chu's definition for spherical mode $Q_n^{\text{TE+TM}}$ is used, this maximum $D/Q$ ratio seems to provide a useful bound for $ka > 1$. For example, when $ka = 5$ the maximum $D/Q_{\text{Chu}} = 35$, which suggests the maximum antenna directivity must be less than 15.4 dB when $Q_{\text{Chu}} = 1$. For reference, this value of directivity is comparable to that of an antenna with $ka = 5$ and 100% aperture efficiency ($D_{100\%\,\text{eff}} = 14$ dB). However, we argue that Chu's definition for $Q$ should be updated. When our updated definition of $Q_n^{\text{TE+TM}}$ from (30) is inserted into (32), we find that the $D/Q$ limit imposes a maximum possible antenna directivity of 50 dB when $Q_{\text{Pfeiffer}} = 1$. It is immediately apparent that this directivity bound is not particularly useful because it turns out to be too loose (i.e., overly optimistic) when the updated $Q$ is applied. Armed with the new definition for $Q$, we thus seek to formulate a new way to look at the relationship between antenna size, directivity, and bandwidth. Rather than optimizing the $D/Q$ ratio, we specify a desired $Q$ and maximize directivity, which is a much more practical metric.

As shown in [25], the antenna directivity is maximized when the $n$th order TE$_{n1}$ and TM$_{n1}$ modes are radiated with equal amplitude denoted as $a_n$. By properly phasing each mode so that they add constructively in the far field, the antenna directivity can be simplified as,

$$D(|a_n|) = \frac{(\sum_n |a_n|)^2}{\sum_n |a_n|^2/(2n+1)} \tag{33}$$

while the overall antenna $Q$ is a weighted average of the quality factors of the constituent $n$th-order spherical modes given by (30),

$$Q(|a_n|) = \frac{\sum_n |a_n|^2 Q_n/(2n+1)}{\sum_n |a_n|^2/(2n+1)} \tag{34}$$

The goal then is to simply find the magnitude of the modal coefficients, $|a_n|$, that maximize $D$ for a specified $Q$. This optimization problem is straightforward to solve using the method of Lagrange multipliers [28, 29],

$$\mathcal{L}(|a_n|) = D(|a_n|) - \mu_1 Q(|a_n|) \tag{35}$$

where $\mu_1$ is the Lagrange multiplier (not to be confused with permeability). Differentiating (35) results in,

$$\frac{\mathrm{d}\mathcal{L}}{\mathrm{d}|a_n|} = \frac{\mathrm{d}D}{\mathrm{d}|a_n|} - \mu_1 \frac{\mathrm{d}Q}{\mathrm{d}|a_n|} \tag{36}$$

where,

$$\frac{\mathrm{d}D}{\mathrm{d}|a_n|} = \frac{2\sum_m |a_m|}{\sum_m \frac{|a_m|^2}{2m+1}} - \frac{(\sum_m |a_m|)^2}{\left(\sum_m \frac{|a_m|^2}{2m+1}\right)^2} \left[\frac{2|a_n|}{2n+1}\right] \tag{37}$$

$$\frac{\mathrm{d}Q}{\mathrm{d}|a_n|} = \left[\frac{2|a_n|}{2n+1}\right] \frac{Q_n}{\sum_m \frac{|a_m|^2}{2m+1}} - \frac{\sum_m \frac{|a_m|^2 Q_m}{(2m+1)}}{\left(\sum_m \frac{|a_m|^2}{2m+1}\right)^2} \left[\frac{2|a_n|}{2n+1}\right] \tag{38}$$

The optimal modal coefficients $|a_n|$ are then found by setting $\frac{\mathrm{d}\mathcal{L}}{\mathrm{d}|a_n|} = 0$ and solving for $|a_n|$,

$$|a_n| = \frac{2n+1}{2}\left[\frac{2\sum_m |a_m|}{\sum_m \frac{|a_m|^2}{2m+1}}\right]$$

$$\times \left[\frac{\mu_1 Q_n}{\sum_m \frac{|a_m|^2}{2m+1}} - \frac{\mu_1 \sum_m \frac{|a_m|^2 Q_m}{(2m+1)}}{\left(\sum_m \frac{|a_m|^2}{2m+1}\right)^2} + \frac{(\sum_m |a_m|)^2}{\left(\sum_m \frac{|a_m|^2}{2m+1}\right)^2}\right]^{-1} \tag{39}$$

Eliminating arbitrary constants results in a simple expression for the optimal modal coefficients,

$$|a_n| = \frac{2n+1}{Q_n + \mu} \tag{40}$$

The desired antenna $Q$ and modal coefficients from (40) are inserted into (34), and the Lagrange multiplier, $\mu$, is solved numerically. Solving for $\mu$ is numerically trivial since the antenna $Q$ monotonically increases with $\mu$, which leads to a single variable convex optimization problem. Once $\mu$ is solved, the maximum directivity for the specified $Q$ is calculated by inserting (40) into (33). Alternatively, the same process can calculate the minimum $Q$ for a specified directivity by inserting (40) into (33), solving for $\mu$, and then inserting the optimal modal coefficients into (34). The lowest order mode has the lowest quality factor, $Q_1 < Q_n$ for all $n > 1$. Therefore, $\mu$ must satisfy $\mu \geq -Q_1$ to ensure the magnitudes of all modal coefficients are positive, $|a_n| \geq 0$. Thus, setting $\mu = -Q_1$ results in the minimum possible $Q$ and directivity. Taking the limit $\mu \to \infty$ recovers the modal coefficients that maximize directivity found by Harrington, which provide infinite $Q$ and directivity when the number of modes, $N \to \infty$. Letting $\mu = 0$ results in Geyi's modal coefficients that maximize the $D/Q$ ratio [27]. Letting $\mu = 1$ generates the heuristic expression for the maximum 'practical' directivity that was proposed in [45].

As an illustration, some optimal spherical mode excitations are computed that either maximize directivity for a specified $Q$,

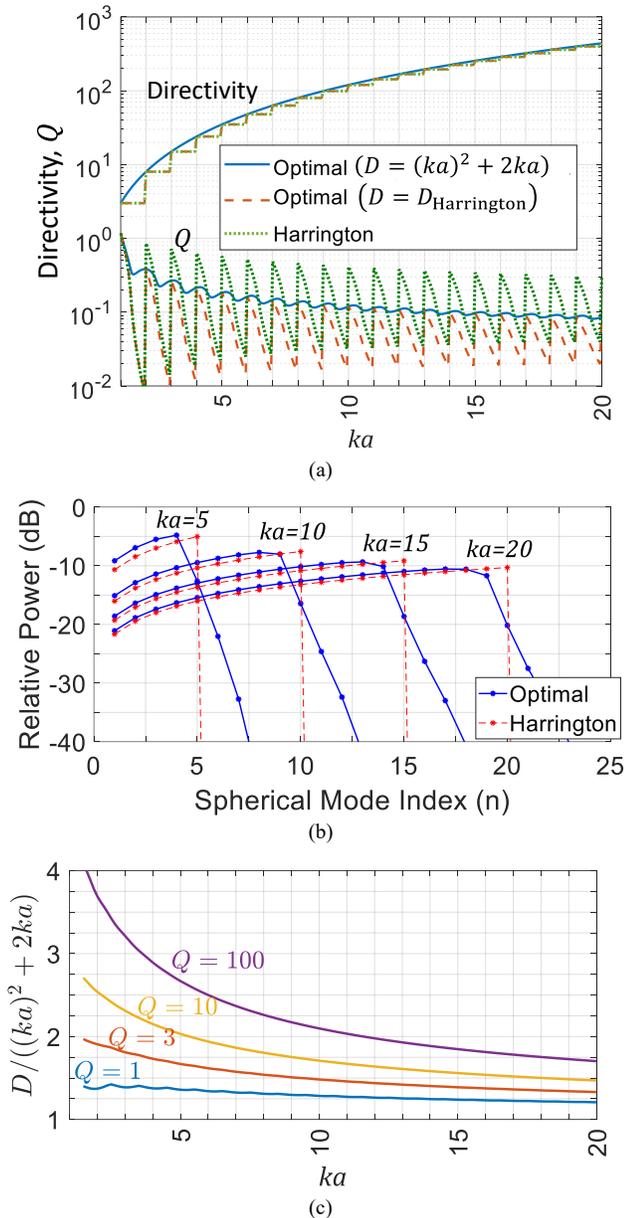

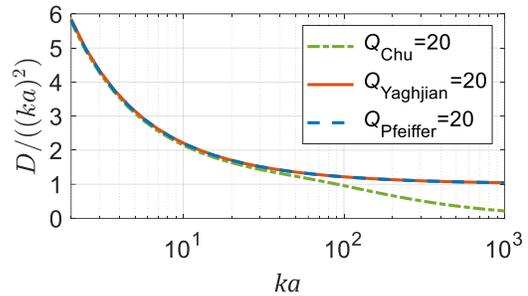

Fig. 13. Maximum aperture efficiency vs. $ka$ when the stipulated overall $Q = 20$. Different curves correspond to using different definitions for $Q_n$ in (40).

truncation $n \leq ka$. Finally, the dashed red curves in Fig. 12(a) correspond to directivity and $Q$ that result from using the optimal modal excitation that achieves the minimum $Q$ with the same equivalent directivity as Harrington's excitation. For this same staircase directivity, the new excitation given by (40) always results in a lower $Q$ than Harrington's excitation.

Fig. 12(b) compares the optimal modal coefficients from (40) to those calculated by Harrington for the cases with antenna sizes $ka = 5, 10, 15,$ and $20$. Our optimal coefficients generate the same directivity as Harrington's analysis ($D = (ka)^2 + 2ka$), but with minimum $Q$ (i.e., the blue and red curves of Fig. 12(a) at the specified $ka$ values). The optimal coefficients tend to excite the lower radiating modes with higher power when compared to Harrington's excitation. Near $n = ka$, the optimal coefficients have a more gradual reduction in power with respect to $n$ compared to the abrupt truncation for Harrington's excitation. Thus, by judiciously employing several modes beyond $n = ka$ via the method of Lagrange multipliers, a lower $Q$ previously not known can be obtained.

Using the new definition of $Q$, Fig. 12(c) plots the maximum antenna directivity vs. $ka$ when $Q = 1, 3, 10,$ and $100$. The directivity is normalized to Harrington's definition of normal directivity. The maximum normalized directivity of the $Q = 1$ curve is within a factor of $1.4\times$ compared to Harrington's directivity (i.e., 1.5 dB) for all values of $ka > 1$. This fact suggests Harrington's simple expression for normal directivity is indeed a decent estimate of the maximum directivity for $Q = 1$. It is also worth noting that the $Q = 3$ curve intersects the point $ka = 6.75$ and $D = 19.7$ dB (i.e., $> 200\%$ aperture efficiency). This means that the effective area of an antenna can be twice as large as the projected area while still achieving a moderate directivity and bandwidth, which is a promising fact for super-gain research. This is once again due to the use of modes with $n > ka$, which provides a measurable improvement for intermediately sized antennas.

Fig. 13 plots the maximum possible aperture efficiency when $Q = 20$ at larger values of $ka$. The different curves correspond to using different definitions for $Q_n$ in (40). The new field-based $Q_{\text{Pfeiffer}}$ closely matches the impedance-based $Q_{\text{Yaghjian}}$. The maximum aperture efficiency is always above unity which agrees with intuition. In contrast, [28, 29] use Chu's definition for $Q_n$, which gives in an unphysical result that suggests the maximum aperture efficiency asymptotically approaches 0 as $ka$ increases. This is another example showing that Chu's or minimize $Q$ for a specified directivity. The solid blue curve in Fig. 12(a) plots the minimum antenna $Q$ vs. $ka$ when the directivity equals the 'normal directivity' defined by Harrington, $D = (ka)^2 + 2ka$. Harrington defined the normal directivity to represent the maximum directivity such that the antenna $Q$ is not 'large'. Our analysis makes the definition explicit by quantifying the minimum $Q$ for this specified directivity. The dotted green curves correspond to the directivity and $Q$ that results from using Harrington's modal coefficients, $|a_n| = 2n + 1$ for $n \leq ka$. The dotted green curves jump when $ka$ is an integer due to Harrington's definition for $Q_n$ can be problematic when applied to larger antennas since commonplace dish antennas with directivities > 40 dB, 50% aperture efficiency, and operating over a waveguide band surpass the derived upper bound on directivity and/or bandwidth.

Fig. 12. (a) Directivity and $Q$ for different spherical mode excitations. Optimal curves correspond to the excitation that minimizes $Q$ for the specified directivity. Harrington curves corresponds to the optimal excitation when the number of spherical harmonics, $N$, is truncated to $N \leq ka$. (b) Comparison of modal coefficients between Harrington's excitation and the optimal excitation that achieves minimum $Q$ and the same directivity as Harrington's excitation. (c) Maximum directivity normalized by normal directivity as a function of antenna electrical size, $ka$, and specified $Q$.

The optimal excitation that radiates modes with $n > ka$ risks reducing the radiation efficiency for practical antennas that have finite conductivity [25, 2, 5]. Thus, the impact of material loss could significantly influence the gain-bandwidth tradeoff that is analyzed here. A relevant data point is to compare our results to [2, 5], which calculates the maximum possible gain from a spherical wire antenna that is limited only by material loss (i.e., no bandwidth limitation). For example, when $ka = 6.75$ and the surface resistance is 0.01 Ω/□ (i.e., copper at 1.5 GHz), [2] calculates a maximum achievable gain of 21.6 dB (i.e., 320% aperture efficiency). Our analysis suggests that the minimum possible $Q$ equals 130 when $ka = 6.75$ and $D = 21.6$ dB, which is quite narrowband. This is an example where bandwidth limitations might be more of a concern than material losses, depending upon the application.

## VI. Conclusion

Limitations of previous definitions of antenna $Q$ are discussed. It is shown that the well-known Chu limit for spherical mode propagation can dramatically underestimate the impedance bandwidth, which motivates an updated analysis. Spherical mode radiation is reexamined, and a new transmission line model is derived that exactly models the fields. The novel transmission line model demonstrates that spherical mode radiation is analogous to wave propagation through a tapered waveguide with cutoff frequency $kr = \sqrt{n(n+1)}$. At smaller radii the waveguide is below cutoff and the fields don't propagate, but instead evanescently grow/decay. At larger radii, the waveguide supports propagating modes in the forward and backward directions. In contrast to Collin and Rothschild's analysis [9] as well as many works [15, 30, 31, 27], our model accounts for the fact that electromagnetic energy generally propagates slower than the speed of light near the antenna when the fields are not TEM polarized, and the energy velocity asymptotically approaches the speed of light in the limit $r \to \infty$. Transforming radiation into a transmission line problem provides a precise and intuitive definition for stored and radiated energy, which in turn leads to a new definition of $Q$ that is valid for arbitrary spherical mode orders and electrical size. The updated energy-based $Q$ agrees with the achievable bandwidth for the wide range of relevant scenarios that are considered here (e.g., small/large $ka$ and $n$, non-resonant, self-resonant, waveguides). This is conceptually attractive since $Q$ is typically defined in terms of stored and dissipated energy. In contrast, $Q_{\text{Yaghjian}}$ is a popular complementary definition of $Q$ that circumvents the notion of stored energy and is instead defined in terms of impedance bandwidth itself [30].

Next, antenna directivity and bandwidth bounds are considered. The method of Lagrange multipliers is used to calculate the first ever definition of the maximum achievable directivity for a specified $Q$ that is valid for arbitrarily large radiators, which reconciles previous bounds as special cases. This analysis provides further evidence that $Q_{\text{Chu}}$ needs updating since the maximum aperture efficiency asymptotically approaches 0 as $ka$ increases when Chu's definition for $Q$ is employed. In contrast, the updated $Q_{\text{Pfeiffer}}$ definition introduced here behaves as expected with the maximum aperture efficiency approaching unity as $ka$ increases.

A natural extension of this work is to consider extending the analysis to non-spherical geometries to provide a tighter bound for arbitrarily shaped antennas that could also include internal stored energy. These updated bounds could be used as a ground truth to compare to more general definitions of $Q$ [21] such as those relying on the notion of recoverable energy [52, 53] or Brune circuit synthesis [16].

## Acknowledgement

This work was performed in support of the EQuAL-P program of the Office of the Director of National Intelligence (ODNI), Intelligence Advanced Research Projects Activity (IARPA).

## Appendix A
### Discussion on Various Velocity Definitions

The radiative energy velocity defined in (23) complements the more conventional waveguide group velocity and total energy velocity. The group velocity, $v_g$, is the speed that a narrowband pulse propagates along a waveguide and is defined as,

$$v_g = \left(\frac{d\beta(\omega)}{d\omega}\right)^{-1} \quad (41)$$

The total energy velocity, $v_e$, is the speed at which the total electromagnetic energy (stored plus radiative) moves along a waveguide and is typically defined as [54],

$$v_e = \frac{P}{\iint_S U \, dS} = \frac{1/2 \iint_S \text{Re}(\bar{E} \times \bar{H}^*) \cdot \hat{n} \, dS}{\iint_S U \, dS} \quad (42)$$

where $S$ corresponds to the waveguide cross section and $\hat{n}$ is the unit vector normal to the surface. Intuitively, the radiative energy velocity, group velocity and total energy velocity are all equal when there is only a forward propagating wave. However, this is not true when there are two counter-propagating waveguide modes. For example, the three differently defined velocities vs. radial position on the spherical waveguide for the $n = 1$ mode are plotted in Fig. 14. The three different velocities diverge near the waveguide cutoff frequency when $|\Gamma| = 1$. However, when $kr > 2$ the reflection coefficient is vanishingly small ($|\Gamma|^2 < 0.01$) and all three velocities converge to a value that is less than the speed of light.

When the radiative energy propagates at less than the speed of light, the total energy density exceeds that of a plane wave in free space carrying the same amount of power. Previous analyses interpreted this increased energy density as stored energy, which in turn resulted in an increased $Q$. There is a small window near cutoff where the radiative energy velocity asymptotically approaches infinity. This creates a continuous transition in stored energy above and below cutoff since $U_r(r) = 0$, when radiative energy velocity is infinite ($v_r \to \infty$). While the radiative energy velocity is infinite at cutoff, the total energy velocity ($v_e$) remains less than the speed of light.

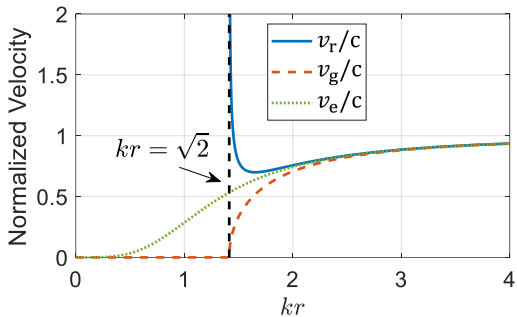

Fig. 14. Energy velocities vs. radial position of the TM$_{1m}$ mode.

These velocity definitions can readily be applied to arbitrary waveguides. Therefore, let us also consider a conceptually simpler scenario of the lowest order TE$_{10}$ mode within a uniform rectangular metallic waveguide to gain physical insight (see Fig. 15(a)). The propagation constant of the rectangular waveguide ($\beta_{rec}$) is defined by the waveguide dimensions,

$$\beta_{rec}(\omega) = \sqrt{k^2 - (\pi/a)^2} \qquad (43)$$

This propagation constant has the same frequency dependence as the spherical waveguide in (10). Again, let $\Gamma$ be the ratio of the tangential electric field of waves propagating in the $-z$ direction and $+z$ directions. The TE$_{10}$ mode fields are inserted into (23), (41), and (42) to calculate the three different velocities. Fig. 15(b) compares the velocities vs. longitudinal position of the waveguide for the case where $|\Gamma|^2 = 0.5$ and $a = \sqrt{2}\lambda$. As expected, the group velocity depends only on the waveguide size and does not vary with longitudinal position. However, the total energy velocity and radiative energy are spatially dependent on $z$ (or equivalently the phase of $\Gamma$). In other words, electromagnetic energy must speed up and slow down to maintain a net power flow at a constant rate along the waveguide.

Fig. 15 (c) and (d) plot the total energy velocity and radiative energy velocity vs. waveguide group velocity for a few different values of the reflection coefficient. As expected, the velocities are all equal when $\Gamma = 0$. The velocities are also physically intuitive when $v_g = c$ (i.e., TEM mode) since the radiative energy velocity equals the group velocity and the total energy velocity is weighted by the percentage of energy flowing in the forward direction. The total energy velocity is always less than the speed of light, but the radiative energy velocity can be much larger depending on the group velocity and reflection coefficient. For example, near cutoff ($v_g \approx 0$), when $\Gamma \to -1$, with a fixed net power flow ($P^+ - P^-$), the sum of the power flowing in the forward and backward directions ($P^+ + P^-$) grows while the total energy approaches zero. Thus, (23) suggests the radiative energy velocity should grow to infinity. This result is intuitive because a high energy velocity is associated with a large power flow and low energy density. While total energy velocity is proportional to the net power flow 'through' the waveguide cross section ($P^+ - P^-$), the radiative energy velocity is proportional to the total power 'into' the waveguide cross section ($P^+ + P^-$).

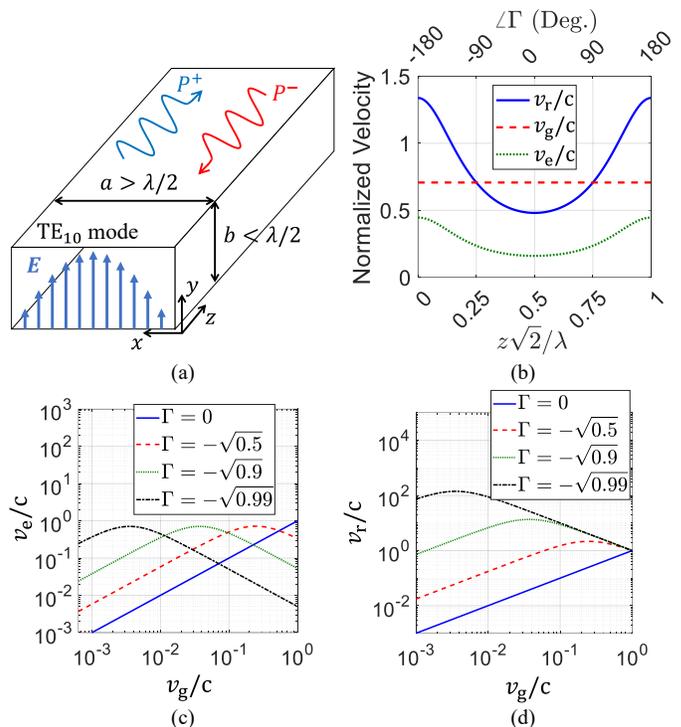

Fig. 15. (a) TE$_{10}$ mode in a rectangular waveguide above cutoff. (b) Different energy velocity definitions vs. longitudinal position on the waveguide for the case where $|\Gamma|^2 = 0.5$ and $a = \sqrt{2}\lambda$. (c) Total energy velocity vs. group velocity for different reflection coefficients. (d) Radiative energy velocity vs. group velocity for different reflection coefficients.

It is also interesting that $v_e$, $v_r$, and $v_g$ have a physically intuitive relationship when they are averaged over a spatial period ($\pi/\beta_{rec}$). More precisely, it is straightforward to show that the TE$_{10}$ mode of a rectangular waveguide satisfies the following conditions for arbitrary waveguide size ($a$) and reflection coefficient ($\Gamma$),

$$\frac{\pi}{\beta_{rec}} \left[ \int_0^{\frac{\pi}{\beta_{rec}}} \frac{1}{v_r(z)} dz \right]^{-1} = v_g \qquad (44)$$

$$\frac{\pi}{\beta_{rec}} \left[ \int_0^{\frac{\pi}{\beta_{rec}}} \frac{1}{v_e(z)} dz \right]^{-1} = \left( \frac{1 - |\Gamma(r)|^2}{1 + |\Gamma(r)|^2} \right) v_g \qquad (45)$$

We suspect this relationship is true for arbitrary waveguides and modes but have not proved it. The integrals on the left-hand side of (44) and (45) represent the time that it takes radiative and total energy to propagate a distance $\pi/\beta_{rec}$, respectively [55]. Therefore, the average velocity is calculated by taking the ratio of this distance to time. The averaged velocities agree with the intuition that radiative energy should propagate at the group velocity, while the total energy velocity equals a weighted average of the stored energy velocity (i.e., 0 m/s) and the radiative energy velocity ($v_r$).

## Appendix B
### Energy Density of Outward And Inward Propagating Waves

One of the key results is that the electric energy density of the outward propagating spherical wave is identical to the magnetic energy density. Consider any conventional waveguides above cutoff. The energy density (J/m) in the differential volume between two adjacent arbitrary cross sections (e.g., within a spherical shell in our case) of the outward propagating magnetic field ($U_{\text{shell}}^{\text{m+}}$) is identical to that of the outward propagating electric field ($U_{\text{shell}}^{\text{e+}}$). To illustrate that for spherical waves, (18) can be derived by substituting in all previously defined quantities. For simplicity, we will consider spherical modes with $m = 0$ here, but it can be readily verified that the result also holds for $m \neq 0$.

From (16), the outward propagating magnetic field is related to the total magnetic field by,

$$U_{\text{shell}}^{\text{m+}} = \frac{\mu_0}{4}\int_{4\pi}|H_\phi^+|^2 r^2 d\Omega = \frac{\mu_0}{4}\int_{4\pi}\left|\frac{H_\phi}{1-\Gamma}\right|^2 r^2 d\Omega \quad (46)$$

Since Substituting $H_\phi$ with the magnetic field for a $\text{TM}_{n,0}$ mode from (3),

$$U_{\text{shell}}^{\text{m+}} = \frac{\mu_0}{4}\left|\frac{C_n[krh_n^{(2)}(kr)]}{1-\Gamma}\right|^2 2\pi\int_0^\pi \left|\frac{dP_n(\cos\theta)}{d\theta}\right|^2 \sin(\theta)\, d\theta \quad (47)$$

Upon integration,

$$U_{\text{shell}}^{\text{m+}} = \frac{\mu_0}{4}\left|\frac{C_n[krh_n^{(2)}(kr)]}{1-\Gamma}\right|^2 \frac{4\pi n(n+1)}{2n+1} \quad (48)$$

Above cutoff, we can substitute in the definition of the reflection coefficient, $\Gamma$, from (14),

$$U_{\text{shell}}^{\text{m+}} = \frac{\mu_0}{4}\left|\frac{C_n[krh_n^{(2)}(kr)](\eta_n^{\text{TM}}+Z_0^{\text{TM}})}{2Z_0^{\text{TM}}}\right|^2 \frac{4\pi n(n+1)}{2n+1} \quad (49)$$

Inserting the characteristic impedance from (11) into the denominator and simplifying results in the magnetic energy density of an outward propagating wave,

$$U_{\text{shell}}^{\text{m+}} = \frac{k}{\eta_0 \omega}\left|C_n[krh_n^{(2)}(kr)](\eta_n^{\text{TM}}+Z_0^{\text{TM}})\right|^2 \\ \times \frac{\pi n(n+1)}{4(2n+1)\left(1 - \frac{n(n+1)}{(kr)^2}\right)} \quad (50)$$

Likewise, the energy density of the outward propagating electric field $E_{\theta,r}^+$ is related to the total field $E_{\theta,r}$ by (16) and (17),

$$U_{\text{shell}}^{\text{e+}} = \frac{\varepsilon_0}{4}\int_{4\pi}(|E_\theta^+|^2 + |E_r^+|^2)\, r^2 d\Omega \\ = \frac{\varepsilon_0}{4}\int_{4\pi}\left(\left|\frac{E_\theta}{1+\Gamma}\right|^2 + \left|\frac{E_r}{1-\Gamma}\right|^2\right) r^2 d\Omega \quad (51)$$

Substituting in the definition of the reflection coefficient above cutoff from (14),

$$U_{\text{shell}}^{\text{e+}} \\ = \frac{\varepsilon_0}{4}\left|\frac{C_n j\eta_0\left(\frac{d[krh_n^{(2)}(kr)]}{dr}\right)(\eta_n^{\text{TM}}+Z_0^{\text{TM}})}{2kr\eta_n^{\text{TM}}}\right|^2 \\ \times 2\pi r^2 \int_0^\pi \left|\frac{dP_n(\cos\theta)}{d\theta}\right|^2 \sin(\theta)\, d\theta \quad (52) \\ + \frac{\varepsilon_0}{4}\left|\frac{C_n j\eta_0 n(n+1)[krh_n^{(2)}(kr)](\eta_n^{\text{TM}}+Z_0^{\text{TM}})}{2kr^2 Z_0^{\text{TM}}}\right|^2 \\ \times 2\pi r^2 \int_0^\pi (P_n(\cos\theta))^2 \sin(\theta)\, d\theta$$

The wave impedance of the outward propagating spherical TM mode ($\eta_n^{\text{TM}}$) from (15) and the characteristic impedance from (11) are inserted into the denominators. In addition, the Legendre polynomial integration is carried out,

$$U_{\text{shell}}^{\text{e+}} = \frac{\varepsilon_0}{4}\left|\frac{C_n \eta_0 [krh_n^{(2)}(kr)](\eta_n^{\text{TM}}+Z_0^{\text{TM}})}{2\eta_0}\right|^2 \\ \times \frac{4\pi n(n+1)}{2n+1}\left[1 + \frac{n(n+1)}{(kr)^2\left(1 - \frac{n(n+1)}{(kr)^2}\right)}\right] \quad (53)$$

The result is then simplified,

$$U_{\text{shell}}^{\text{e+}} = \frac{k}{\eta_0 \omega}\left|C_n[krh_n^{(2)}(kr)](\eta_n^{\text{TM}}+Z_0^{\text{TM}})\right|^2 \\ \times \frac{\pi n(n+1)}{4(2n+1)\left(1 - \frac{n(n+1)}{(kr)^2}\right)} \quad (54)$$

Thus, we show the electric energy density of the outward propagating spherical wave (54) is identical to the magnetic energy density in (50).

It is straightforward to show that the electric and magnetic energy density of inward propagating waves are also equal. From (16), the inward propagating magnetic field is related to the total magnetic field by,

$$U_{\text{shell}}^{\text{m-}} = \frac{\mu_0}{4}\int_{4\pi}|H_\phi^-|^2\, r^2 d\Omega \\ = \frac{\mu_0}{4}\int_{4\pi}\left|\frac{H_\phi}{1/\Gamma - 1}\right|^2 r^2 d\Omega \quad (55) \\ = U_{\text{shell}}^{\text{m+}}|\Gamma|^2$$

The energy density of the inward propagating electric field $E_{\theta,r}^-$ is related to the total field $E_{\theta,r}$ by (16) and (17),

$$U_{\text{shell}}^{\text{e-}} = \frac{\varepsilon_0}{4}\int_{4\pi}(|E_\theta^-|^2 + |E_r^-|^2)\, r^2 d\Omega \\ = \frac{\varepsilon_0}{4}\int_{4\pi}\left(\left|\frac{E_\theta}{1/\Gamma + 1}\right|^2 + \left|\frac{E_r}{1/\Gamma - 1}\right|^2\right) r^2 d\Omega \quad (56) \\ = U_{\text{shell}}^{\text{e+}}|\Gamma|^2$$

which is equal to (55) since $U_{\text{shell}}^{\text{e+}} = U_{\text{shell}}^{\text{m+}}$.